\RequirePackage{fix-cm}
\documentclass[twocolumn]{svjour3}          
\smartqed  
\usepackage{graphicx}
\usepackage{color}
\begin{document}

\title{Peculiarities of length scales in a two-orbital superconductor}



\author{Grzegorz Litak \and Teet \"{O}rd         \and
        K\"{u}llike R\"{a}go      \and
        Artjom Vargunin         
}


\institute{
Grzegorz Litak \at Faculty of Mechanical Engineering, Lublin University of Technology, Nadbystrzycka 36,
PL-20618 Lublin, Poland \\ \email{g.litak@pollub.pl} \\ \\
Teet \"{O}rd, K\"{u}llike R\"{a}go,  Artjom Vargunin \at
            Institute of Physics, University of Tartu \\ T\"{a}he 4, 51010 Tartu, Estonia \\
               \email{teet.ord@ut.ee}          
}

\date{Received: date / Accepted: date}

\maketitle

\begin{abstract}
We study the spatial behaviour of coherency and magnetic field in a two-orbital superconductor. The superconducting phase transition is caused here by the on-site intra-orbital attractions (negative-$U$ Hubbard model) and inter-orbital pair-tran\-sfer interaction. We find the critical (diverging at $T_{c}$) and non-critical (remaining finite) coherence lengths and magnetic field penetration depth for various values of hopping integrals and the strengths of intra-orbital attractions. Numerical results have been obtained for a two-dimensional square lattice.
\end{abstract}

\section{Introduction}

The multi-component structure of electron system plays the crucial role in the formation of superconducting state in a number of novel compounds, see \cite{bianconi1,bianconi2,aoki,bianconi3}. For the present time the multi-band nature of superconductivity has been identified, among the others, in MgB$_2$ \cite{MgB2}, cuprates \cite{cuprates}, iron-arsenic compounds \cite{pnictides}, V$_3$Si \cite{yu2005}, NbSe$_{2}$ \cite{NbSe2}, and Sr$_2$RuO$_4$ \cite{maeno2001,annett2003}. The presence of interacting superconductivity order parameters is decisive for these systems.

Recent studies \cite{babaev,babaev1,babaev2,shanenko1,babaev3,ord1,litak2012,shanenko2} have revealed unexpected properties of two-band inhomogeneous superconductivity compared to one-band case. Interband pair-transfer interaction modifies qualitatively the coherence lengths in two-band scenario \cite{babaev1,babaev2,babaev3,ord1}. In this respect there is strong necessity of reinterpretation of relevant experimental data. Moreover, the character of spatial coherency affects several theoretical conception, e.g. type-1.5 superconductivity \cite{babaev1} suggested to explain novel patterns in vortex structure in magnesium diboride \cite{moshchalkov} and strontium ruthenate \cite{babaev4}.

The characteristic lengths of a two-orbital superconductor described by negative-$U$ Hubbard model \cite{Ref mrr} were previously analyzed in dependence on band filling for various inter-orbital interactions and orbital energies \cite{litak2012,litak2012a}. As a continuation of that research we present below the results of the model calculations for the superconducting state and related length scales with varying strengths of intra-orbital interactions and hopping integrals.

\section{Basic equations}

The bulk superconducting state in a two-orbital system is described by the Hartree-Fock-Gorkov self-consistent equations
\begin{eqnarray}\label{e1}
\Delta_{\alpha}=\frac{-1}{N}\sum_{\alpha'\mathbf{k}}U^{\alpha\alpha'}\frac{\Delta_{\alpha}}{2E_{\alpha}(\mathbf{k})}
\tanh\frac{E_{\alpha}(\mathbf{k})}{2k_\mathrm{B}T} \, ,
\end{eqnarray}
\begin{eqnarray}\label{e2}
n_{\alpha}=\frac{1}{2}\left[1
-\frac{\tilde{\varepsilon}_{\alpha}(\mathbf{k})}{E_{\alpha}(\mathbf{k})}\tanh\frac{E_{\alpha}(\mathbf{k})}{2k_{B}T}\right] \, .
\end{eqnarray}
Here $\alpha=1,2$ is the orbital index; $N$ is the number of lattice sites; $\Delta_{\alpha}$ are the superconductivity gaps; $n_{\alpha}$ are the occupation numbers of orbitals per lattice site;  $U^{\alpha\alpha}<0$ are the on-site intra-orbital attraction energies and $U^{12}=U^{21}$ is the inter-orbital interaction energy. Note that both intra- and interorbital channels support superconductivity. Futher, $E_{\alpha}(\mathbf{k})=\sqrt{\tilde{\varepsilon}^{2}_{\alpha}(\mathbf{k})+\left|\Delta_{\alpha}\right|^{2}}$, where $\tilde{\varepsilon}_{\alpha}(\mathbf{k})=\varepsilon_{\alpha}(\mathbf{k})+\frac{1}{2}U^{\alpha\alpha}n_{\alpha}-\mu$ and $\varepsilon_{\alpha}(\mathbf{k})$ is the electron band energy associated with the orbital $\alpha$; $\mu$ is the chemical potential determined by the equation $n_{1}+n_{2}=n$ for the total number of electrons per site $n$. The equations (\ref{e1}) and (\ref{e2}) follow in the mean-field approximation (see \cite{litak2012}) from the Hamiltonian of the two-orbital negative-$U$ Hubbard model
\begin{eqnarray}\label{e3}
H&=&\sum_{\alpha}\sum_{i,j}\sum_{\sigma}\left[t_{ij}^{\alpha\alpha}+\left(\varepsilon_{\alpha}^{0}-\mu\right)
\delta_{ij}\right]a_{i\alpha\sigma}^{+}a_{j\alpha\sigma} \nonumber \\
&+&\frac{1}{2}\sum_{\alpha}\sum_{i}\sum_{\sigma}U^{\alpha\alpha} n_{i\alpha\sigma}n_{i\alpha-\sigma}\nonumber \\
&+&\frac{1}{2}\sum_{\alpha\neq\alpha'}\sum_{i}\sum_{\sigma}U^{\alpha\alpha'}a_{i\alpha\sigma}^{+}a_{i\alpha'\sigma}
a_{i\alpha-\sigma}^{+}a_{i\alpha'-\sigma} \, ,
\end{eqnarray}
where $a_{i\alpha\sigma}^{+}$ is the electron creation operator in the orbital $\alpha=1,2$ localized at the site $i$; $\sigma$ is the spin index; $t_{ij}^{\alpha\alpha}$ is the hopping integral and $\varepsilon_{\alpha}^{0}$ is the orbital energy. Eq. (\ref{e1}) defines the superconductivity phase transition temperature $T_{c}$ by the equation
\begin{eqnarray}\label{e4}
\left(1+U^{11}g_{1}\left(T_{c}\right)\right)\left(1+U^{22}g_{2}\left(T_{c}\right)\right) \nonumber \\
-\left(U^{12}\right)^{2}g_{1}\left(T_{c}\right)g_{2}\left(T_{c}\right)=0 \, ,
\end{eqnarray}
where
\begin{eqnarray}\label{e5}
g_{\alpha}(T)= \frac{1}{2N}\sum_{\mathbf{k}}\frac{1}{\tilde{\varepsilon}_{\alpha}(\mathbf{k})}
\tanh\frac{\tilde{\varepsilon}_{\alpha}(\mathbf{k})}{2k_{B}T}  \, .
\end{eqnarray}

For the spatially inhomogeneous two-orbital superconductivity the Ginzburg-Landau equations \cite{litak2012,litak2012a} read as
\begin{eqnarray}\label{e6}
\Delta_{\alpha }(\mathbf{r})&=&-\sum_{\alpha '}U^{\alpha\alpha'}\biggl[g_{\alpha'}(T)-\nu_{\alpha'}
\left|\Delta_{\alpha '}(\mathbf{r})\right|^{2} \nonumber \\
&+&\beta_{\alpha '}\left(\nabla +i\frac{2\pi}{\Phi_{0}}\mathbf{A}\right)^{2}\biggr]\Delta_{\alpha '}(\mathbf{r})
\end{eqnarray}
with
\begin{eqnarray}\label{e7}
\nu_{\alpha}
 &=& \frac{-1}{2N}\sum_{\mathbf{k}}\frac{\partial}{\partial
\left|\Delta_{\alpha}\right|^{2}}\left[\frac{1}{E_{\alpha}(\mathbf{k})} \right. \nonumber \\
&\times & \left.\left. \tanh\frac{E_{\alpha}(\mathbf{k})}{2k_{B}T_{c}}\right]\right|_{\Delta_{\alpha}=0}  \, ,
\end{eqnarray}
and
\begin{eqnarray} \label{e8}
\mathbf{j}_{s}&=&-i\frac{2\pi}{V_{0}\Phi_{0}}
\sum_{\alpha}\beta_{\alpha}\left\{\Delta^{\ast}_{\alpha}(\mathbf{r})\nabla\Delta_{\alpha}(\mathbf{r})\right.\nonumber \\
&-& \left.\Delta_{\alpha}(\mathbf{r})\nabla\Delta^{\ast}_{\alpha}(\mathbf{r})\right\}\nonumber \\
&-&\frac{2}{V_{0}}\left(\frac{2\pi}{\Phi_{0}}\right)^{2}
\left(\sum_{\alpha}\beta_{\alpha}|\Delta_{\alpha}(\mathbf{r})|^{2}\right)\mathbf{A} \, ,
\end{eqnarray}
where for isotropic electron spectrum
\begin{eqnarray}\label{e9}
\beta_{\alpha}=\beta_{\alpha 1}=\beta_{\alpha 2}=\beta_{\alpha 3}
\end{eqnarray}
and
\begin{eqnarray}\label{e10}
\beta_{\alpha l}&=& \frac{-1}{4N}\sum_{\mathbf{k}}
\frac{\partial^{2}}{\partial q_{l}^{2}}\left\{
\frac{1}{\tilde{\varepsilon}_{\alpha}(\mathbf{k})+\tilde{\varepsilon}_{\alpha}(\mathbf{k}-\mathbf{q})}\right . \nonumber \\
 &\times& \left[\tanh\left(\frac{\tilde{\varepsilon}_{\alpha}(\mathbf{k})}{2k_{B}T_{c}}\right) \right. \nonumber \\
 &+& \left. \left.\left. \tanh\left(\frac{\tilde{\varepsilon}_{\alpha}(\mathbf{k}-\mathbf{q})}{2k_{B}T_{c}}\right)\right]\right\}\right|_{\mathbf{q}=0}.
\end{eqnarray}
The expression for the density of supercurrent $\mathbf{j}_{s}$ includes the vector potential $\mathbf{A}$, the magnetic flux quantum $\Phi_{0}$ and the volume of unit cell $V_{0}$.

The solutions of inhomogeneous gap equations (\ref{e6}) are scaled by two, critical or soft (index $s$) and non-critical or rigid (index $r$), characteristic lengths \cite{litak2012}
\begin{eqnarray}\label{e11}
\xi_{s,r}^{2}\left(T\right)=\frac{G\left(T\right)
\pm\sqrt{G^{2}\left(T\right)-4K\left(T\right)\gamma}}{2K\left(T\right)}
\end{eqnarray}
with
\begin{eqnarray}\label{e12}
G\left(T\right)&=&\left(U^{12}\right)^{2}\left[\tilde{g}_{1}(T)\beta_{2}+\tilde{g}_{2}(T)\beta_{1}\right] \nonumber\\ &-&\left[1+U^{11}\tilde{g}_{1}(T)
\right]U^{22}\beta_{2} \nonumber\\
&-&\left[1+U^{22}\tilde{g}_{2}(T)\right]U^{11}\beta_{1} \, ,
\end{eqnarray}
\begin{eqnarray}\label{e13}
K\left(T\right)&=&\left[1+U^{11}\tilde{g}_{1}(T)\right]\left[1+U^{22}\tilde{g}_{2}(T)\right] \nonumber\\
&-&\left(U^{12}\right)^{2}\tilde{g}_{1}(T)\tilde{g}_{2}(T) \, ,
\end{eqnarray}
\begin{eqnarray}\label{e14}
\tilde{g}_{\alpha }(T)=g_{\alpha}(T)-3\nu_{\alpha}\left(\Delta_{\alpha }(T)\right)^{2} \, ,
\end{eqnarray}
and
\begin{eqnarray}\label{e15}
\gamma=\left[U^{11}U^{22}-\left(U^{12}\right)^{2}\right]\beta_{1}\beta_{2} \, .
\end{eqnarray}
Eq. (\ref{e8}) yields for the magnetic field penetration depth
\begin{equation} \label{e16}
\lambda\left(T\right)=\sqrt{\frac{V_{0}\Phi^{2}_{0}}{8\pi^{2}\mu_{0}\sum\limits_{\alpha}\beta_{\alpha}\left(\Delta_{\alpha}\left(T\right)\right)^{2}}} \, ,
\end{equation}
where $\mu_{0}$ is the magnetic permeability of free space. We also consider here real homogeneous gaps $\Delta_{\alpha}\left(T\right)$.

In what follows the numerical calculations have been carried out for two-dimensional square lattice with hopping integrals limited to nearest neighbours
$t^{\alpha\alpha}_{ij}=t_\alpha$ and orbital energies $\varepsilon_{1}^{0}=\varepsilon_{2}^{0}=0$. In this case the electron band energies associated with $s$-orbitals are $\varepsilon_{\alpha}(\mathbf{k}) \linebreak =-2t_{\alpha}\left[\cos (ak_{x}) + \cos (ak_{y})\right]$
where $a$ is the lattice constant and $-\pi/a\leq k_{x,y}\leq \pi/a$. We have also chosen $k_{B}=1$.

\begin{figure}[!h]
\includegraphics[width=30mm,angle=-90]{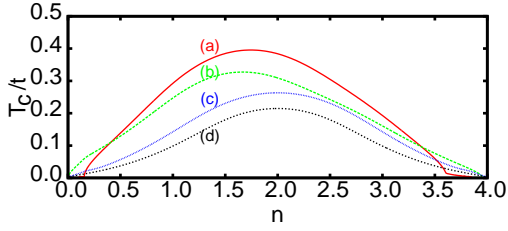}
 \caption{The critical temperature $T_c$ \textit{vs} total number of electrons per lattice site
 $n$ for the chosen set of hopping integrals ($t_1$, $t_2$) and electron-electron interactions ($U^{\alpha\alpha'}$; $\alpha,\alpha'=1,2$);
(a) $t_1= t$, $t_2=0.8t$, $U^{11}=-1.5 t$, $U^{22}=-2.5 t$,
(b) $t_1= 0.8t$, $t_2=t$, $U^{11}=-1.5 t$, $U^{22}=-2.5 t$,
(c) $t_1= 0.8 t$, $t_2=t$, $U^{11}=U^{22}=-2.0 t$,
(d) $t_1= t_2=t$, $U^{11}=U^{22}=-2.0 t$,
respectively, and $|U^{12}|=|U^{21}|=0.04 t$ in each case ($t$ is an arbitrary unit of hopping integral).}
\end{figure}
\begin{figure}[!h]
\vspace{-1.0cm}
\includegraphics[width=50mm,angle=-90]{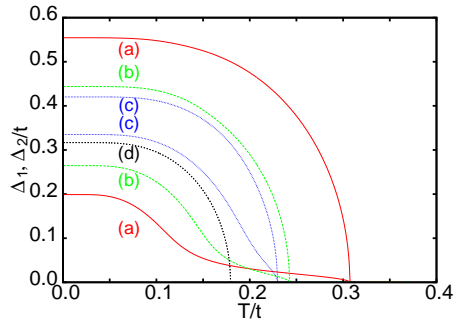}  \\
 \caption{Superconducting gaps $\Delta_1$ and  $\Delta_2$ \textit{vs} temperature for the same sets of parameters as in Fig. 1 with the number of electrons n=2.5. The gaps coincide for the case (d) as the subbands ($\alpha= $ 1 and 2) are equal.}
\end{figure}

\section{Results and discussions}

In Fig. 1 the dependence $T_\mathrm{c}$ \textit{vs} $n$ is shown for various values of $U^{\alpha\alpha}$ and $t_{\alpha}$. The shape of the function $T_\mathrm{c}(n)$ is determined mainly by the following factors: (i) the location of the chemical potential $\mu$ in the separate bands including also Van Hove singularities (intraband contributions to superconductivity) and (ii) the location of $\mu$ in the region of bands overlapping (interband pair-transfer contribution to superconductivity).
Note that the dependence $T_\mathrm{c}(n)$ is symmetric for $t_1=t_2$ and $U^{11}=U^{22}$ (curve (d) in  Fig. 1). By introducing any difference in orbitals the asymmetry of the function $T_\mathrm{c}(n)$ appears. The shift of the $T_\mathrm{c}$ maximum towards the smaller values of $n$ (curves (a), (b), and (c) in Fig. 1) is caused by effective Hartree corrections $U^{\alpha \alpha} n_{\alpha}/2$ and by the difference in particular band-widths ($8t_{\alpha}$) as $U^{11}\neq U^{22}$ and $t_{1}\neq t_{2}$ respectively. The stronger effect of the Hartree renormalization $|\frac{1}{2}U^{11}n_1-\frac{1}{2}U^{22}n_2|$
is supported by the discrepancy of band fillings $n_{1,2}$ which becomes especially large once $\mu$ passes the Van Hove singularity \cite{cw,markiewicz}.

The temperature dependencies of the gaps $\Delta_1$ and $\Delta_2$ (see Fig. 2) illustrate the variation of the driving role of pairing channels in different bands, including the tails (curves (a) and (b)) caused by the interband proximity effect. In the latter case the intra-orbital processes are substantially more dominating in the formation of superconductivity compared to inter-orbital contribution. In particular, there exists a temperature region just below the phase transition point where the pairing in the band with larger gap induces weak superconductivity in the second band via interband pair-transfer of arbitrarily small intensity.

\begin{figure}[!h]
\includegraphics[width=40mm,angle=-90]{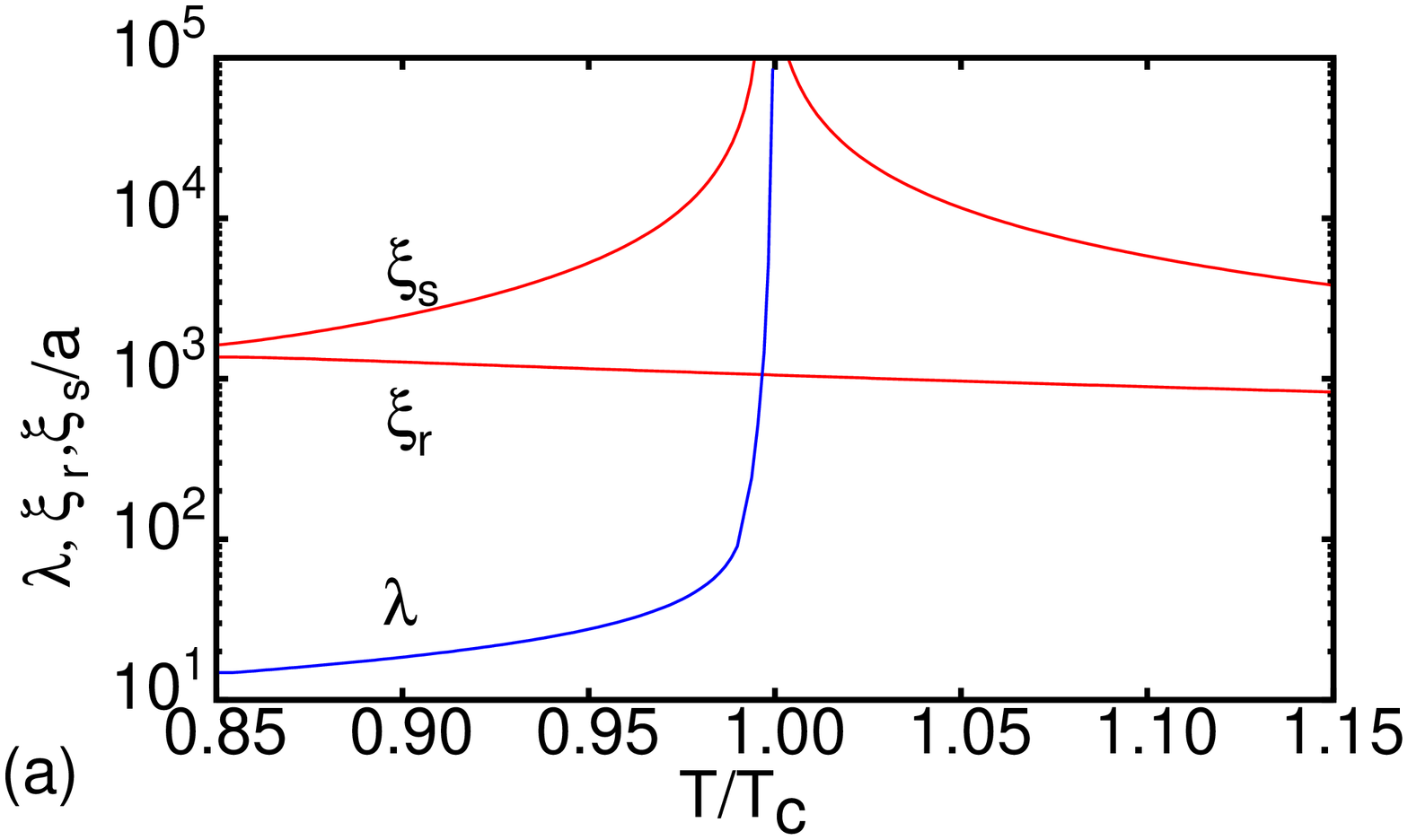}
\includegraphics[width=40mm,angle=-90]{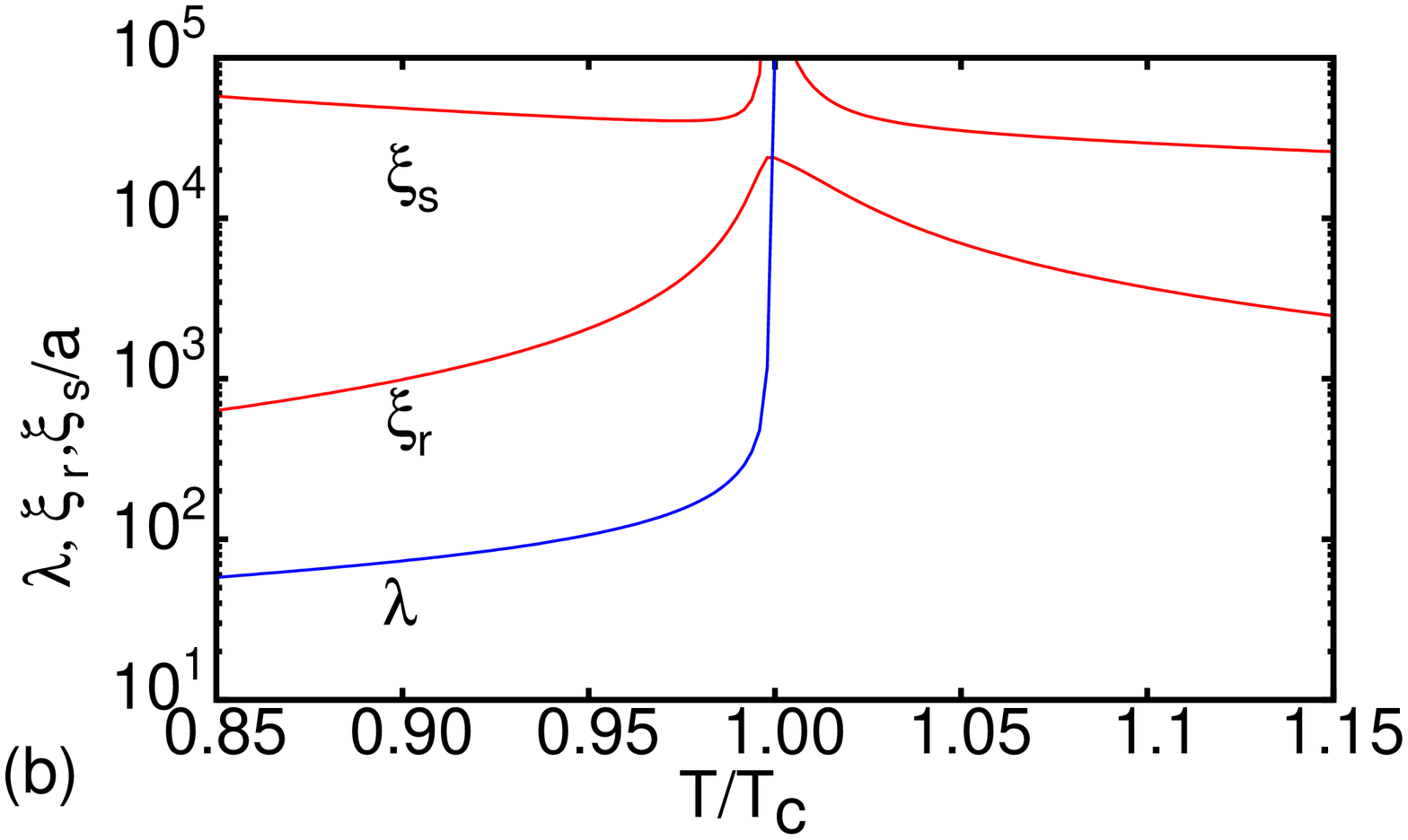}
\includegraphics[width=40mm,angle=-90]{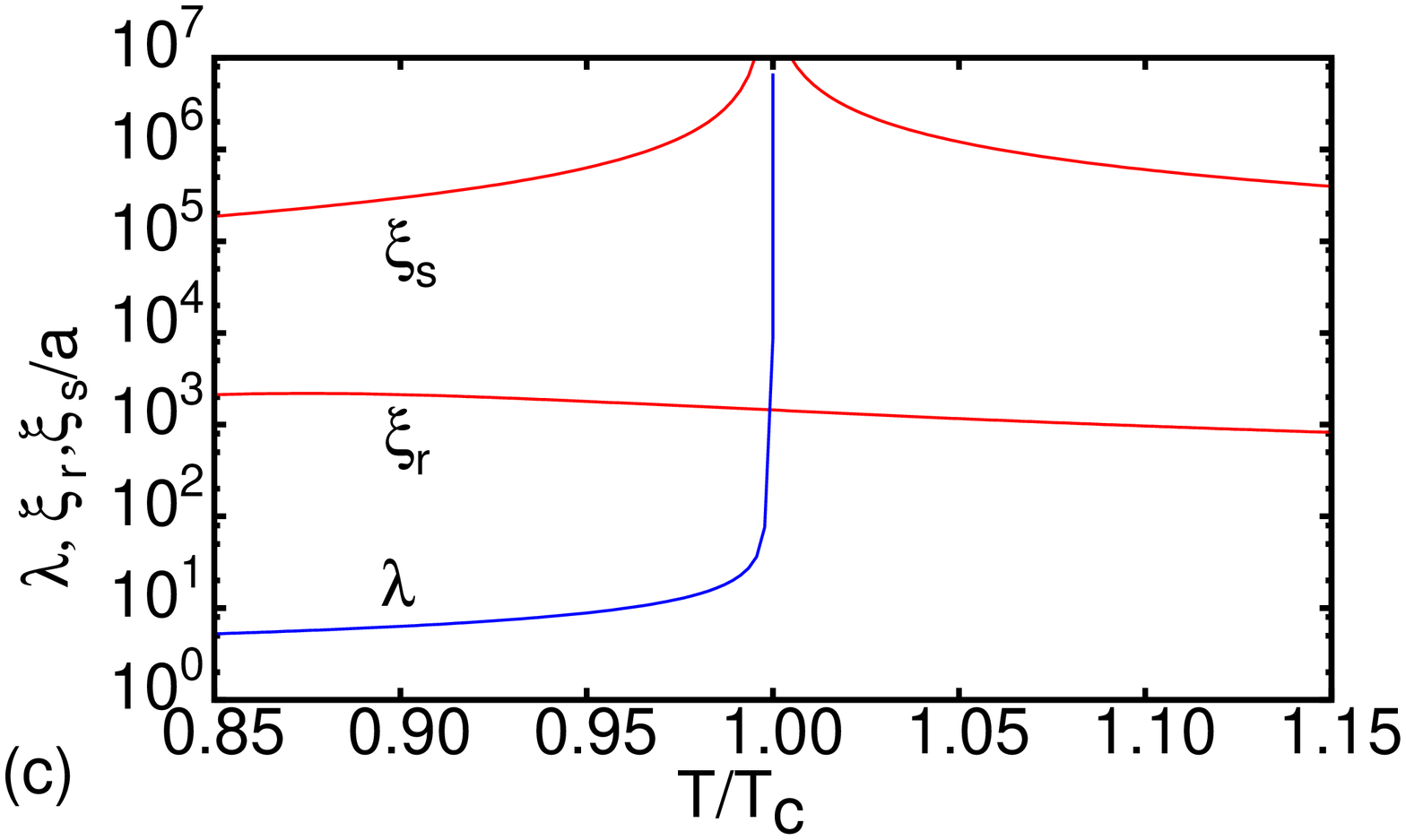}
\includegraphics[width=40mm,angle=-90]{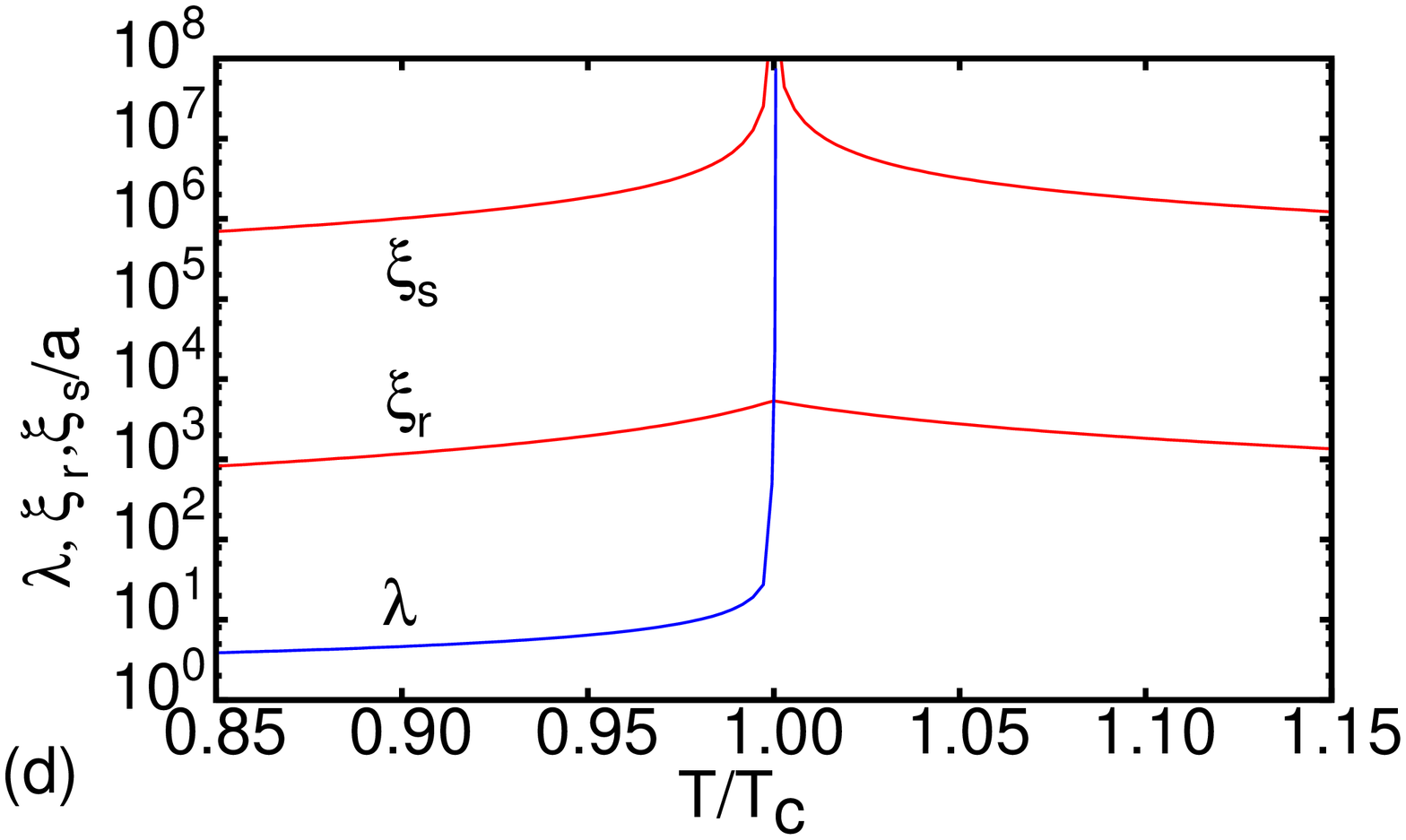}
 \caption{The soft and rigid coherence lengths and penetration depth \textit{vs} temperature. The  parameters as in Fig. 1 with the number of electrons n=2.5.}
\end{figure}

The coherence lengths and magnetic field penetration depth are plotted in Fig. 3. Apart from the divergence of the critical length scale $\xi_{s}$ at $T_\mathrm{c}$, we observe also a maximum for the non-critical length scale $\xi_{r}$ slightly below $T_\mathrm{c}$. This second peak is a combined effect of band filling and interactions in particular bands.

It is worth to mention that for the parameters used there exists always a small
temperature domain in superconducting phase where $\xi_s > \lambda > \xi_r$. The domain width is determined by the values of $\lambda$ and $\xi_r$ which are strongly affected by the interband proximity effect (see also Eq. (\ref{e16})) and discrepancy between intrinsic critical temperatures of the bands. These factors play most in favour in case (a). Note also that the presence of proximity effect can be identified by the non-monotonic temperature dependence of the critical coherence length $\xi_s$ below $T_c$ (c.f. \cite{ord1,litak2012}). The corresponding maximum of $\xi_s(T)$ for $T<T_{c}$ remains out off the scope of panels (a) and (b) in Fig. 3.

\section{Conclusion}

We have examined the temperature behaviour of critical and non-critical coherence lengths, and magnetic field penetration depth in a two-orbital superconductor in dependence on orbital parameters. It was demonstrated that the variation of the combinations of intra-orbital electron-electron interactions and hopping integrals changes substantially the relative efficiency of the pairing channels responsible for the superconducting ordering which is reflected in the peculiarities of the dependencies of characteristic length scales on temperature.

\begin{acknowledgements}
This research was supported by the European Union through the European Regional Development Fund (Centre of Excellence "Mesosystems: Theory and
Applications", TK114). We
acknowledge the support by the Estonian Science Foundation, Grant No 8991. G.L.
kindly acknowledge a financial support by
the European Union 7th Framework
Programme, contract No. FP-7 245479.
\end{acknowledgements}


\begin{thebibliography}{0}

\bibitem{bianconi1}
Bianconi, A.: J. Supercond. Nov. Magn. \textbf{18}, 625 (2005).
%
\bibitem{bianconi2}
Caivano, R., Fratini, M., Poccia, N., Ricci, A., Puri, A., Ren, Z.-A., Dong, X.-L., Yang, J., Lu, W., Zhao, Z.-X., Barba, L., Bianconi, A.: Supercond.
Sci. Technol. \textbf{22}, 014004 (2009).
%
\bibitem{aoki}
Aoki, H.: J. Supercond. Nov. Magn. \textbf{25}, 1243 (2012).
%
\bibitem{bianconi3}
Bianconi, A., Poccia, N.: J. Supercond. Nov. Magn. \textbf{25}, 1403 (2012).
%
\bibitem{MgB2}
%
Tsuda, S., Yokoya, T., Takano, Y., Kito, H., Matsushita, A., Yin, F., Itoh, J., Harima, H., Shin, S.: Phys. Rev. Lett. \textbf{91}, 127001 (2003).
%
\bibitem{cuprates}
Khasanov, R., Shengelaya, A., Maisuradze, A., La Mattina, F., Bussmann-Holder, A., Keller, H., M{\"u}ller, K. A.: Phys. Rev. Lett. \textbf{98}, 057007 (2007).
%
\bibitem{pnictides}
Ding, H., Richard, P., Nakayama, K., Sugawara, K., Arakane, T., Sekiba, Y., Takayama, A., Souma, S., Sato, T., Takahashi, T., Wang, Z., Dai, X., Fang, Z., Chen, G.F., Luo, J.L., Wang,  N.L.: EPL \textbf{83}, 47001 (2008).
%
\bibitem{yu2005} Nefyodov, Yu.A., Shuvaev, A. M., Trunin, M. R.:EPL \textbf{72}, 638 (2005).
%
\bibitem{NbSe2}
Yokoya, T., Kiss, T., Chainani, A., Shin, S., Nohara, M., Takagi, H.: Science \textbf{294}, 2518 (2001).
%
\bibitem{maeno2001} Maeno, Y., Rice, T.M., Sigrist, M.: Phys. Today \textbf{54}, 42 (2001).
%
\bibitem{annett2003} Annett, J.F., Litak, G., Gyorffy, B.L., Wysoki\'nski, K.I., Europ. Phys. J. B \textbf{36}, 301 (2003).
%
%
\bibitem{babaev}
Babaev, E., Carlstr\"{o}m, J., Speight, M.: Phys. Rev. Lett. \textbf{105}, 067003 (2010).
%
\bibitem{babaev1}
Carlstr\"{o}m, J., Babaev, E., Speight, M.: Phys. Rev. B \textbf{83}, 174509 (2011).
%
\bibitem{babaev2}
Silaev, M., Babaev, E.: Phys. Rev. B \textbf{84}, 094515 (2011).
%
%
\bibitem{shanenko1}
Komendova, L., Milosevic, M.V., Shanenko, A.A., Peeters, F.M.: Phys. Rev. B. \textbf{84}, 064522 (2011).
%
\bibitem{babaev3}
Silaev, M., Babaev, E.: Phys. Rev. B \textbf{85}, 134514 (2012).
%
%
\bibitem{ord1}
\"{O}rd, T., R\"{a}go, K., Vargunin, A.: J. Supercond. Novel Magn. \textbf{25}, 1351 (2012).
%
\bibitem{litak2012}
Litak, G., \"{O}rd, T., R\"{a}go, K., Vargunin, A.: Acta Phys. Pol. A \textbf{121}, 747 (2012).
%
%
\bibitem{shanenko2}
Komendova, L., Chen, Y., Shanenko, A.A., Milosevic, M.V., Peeters, F.M.: Phys. Rev. Lett. \textbf{108}, 207002 (2012).
%
\bibitem{moshchalkov}
Moshchalkov, V., Menghini, M., Nishio, T., Chen, Q.H., Silhanek A. V., Dao, V.H., Chibotaru, L.F., Zhigadlo, N.D., Karpinski, J.: Phys. Rev. Lett. \textbf{102}, 117001 (2009).
%
\bibitem{babaev4}
Garaud, J., Agterberg, D.F., Babaev, E.: arXiv: 1207.6395.
%
%
\bibitem{litak2012a}
Litak, G., \"{O}rd, T., R\"{a}go, K., Vargunin, A.: Physica C (2012), doi: http://dx.doi.org/10.1016/j.physc.2012. 06.008 .
%
\bibitem{Ref mrr}
Micnas, R., Ranninger, J., Robaszkiewicz, S.: Rev. Mod. Phys. \textbf{62}, 113 (1990).
%
\bibitem{cw}
Ciechan, A., Wysoki\'nski, K.I.: Phys. Rev. B \textbf{80}, 224523 (2009).
%
\bibitem{markiewicz}
Markiewicz, R. S.: J. Phys. Chem. Solids \textbf{58}, 1179 (1997).
%
\end{thebibliography}
\end{document}